\documentclass[prl,twocolumn,amsmath,amssymb,floatfix,superscriptaddress]{revtex4-1}

\usepackage{graphicx}
\usepackage{amssymb}
\usepackage{amsmath}
\usepackage{color}
\usepackage{ wasysym }
\usepackage{hyperref}

\def\barray{\begin{array}}
\def\earray{\end{array}}
\def\be{\begin{equation}}
\def\ee{\end{equation}}
\def\ben{\begin{equation} \nonumber}
\def\een{\end{equation}}
\def\ban{\begin{eqnarray*}}
\def\ean{\end{eqnarray*}}
\def\ba{\begin{eqnarray}}
\def\ea{\end{eqnarray}}

\def\curv{\mathcal{R}}

\def\({\left(}
\def\){\right)}
\def\[{\left[}
\def\]{\right]}

\def\f{\nu}
\def\nn{\nonumber}

\begin{document}

\title{Leptogenesis from Left-Handed Neutrino Production during Axion Inflation}
\author{Peter Adshead}
\affiliation{Department of Physics, University of Illinois at Urbana-Champaign, Urbana, Illinois 61801, U.S.A.}
\author{Evangelos I. Sfakianakis}
\affiliation{Department of Physics, University of Illinois at Urbana-Champaign, Urbana, Illinois 61801, U.S.A.}

\begin{abstract}
We propose that the observed matter-antimatter asymmetry can be naturally produced as a by-product of axion-driven slow-roll inflation by coupling the axion to by-product neutrinos. We assume that grand unified theory scale right-handed neutrinos are responsible for the masses of the standard model neutrinos and that the Higgs is a light field during inflation and develops a Hubble-scale  {root-mean-square} value. In this setup, the rolling axion generates a helicity asymmetry in by-product neutrinos. Following inflation, this helicity asymmetry becomes equal to a net lepton number as the Higgs  {condensate} decays and is partially reprocessed by the $SU(2)_{L}$ sphaleron into a net baryon number.
\end{abstract}
\maketitle

Axions are attractive candidates for the inflaton because an approximate shift symmetry protects their potential from large radiative corrections~\cite{Freese:1990rb}. However, the simplest models are difficult to realize within UV complete theories because a Planck-scale axion decay constant is required in order to match the amplitude and scale dependence of the observed  curvature fluctuations. Recently, interest in axionic models has been revived with the realization that monodromy effects~\cite{Silverstein:2008sg} can generate a suitable potential for large-field models of inflation. These large-field models may be required if any of the B-mode signal found in the BICEP2 results~\cite{Ade:2014xna} is due to primordial gravitational waves.

Inflation is efficient at accounting for the adiabatic, nearly scale-invariant spectrum of curvature fluctuations in the early Universe with an amplitude measured to be $\Delta_{\curv}^2 \sim 2.2 \times 10^{-9}$ \cite{Ade:2013zuv}. In addition, there is an observed abundance of matter over antimatter, quantified in terms of the baryon-to-photon ratio. Observations of the microwave background constrain the baryon asymmetry parameter to be \cite{Komatsu:2010fb}
\begin{align}\label{eqn:baryonphotonratio}
\eta = \frac{n_b - n_{\bar{b}}}{n_{\gamma}}  = (6.5\pm 0.15)\times10^{-10},
\end{align}
and constraints on nucleosynthesis require  $\eta \sim 5.7-6.7\times10^{-10}$ in order to get the observed light elemental abundances correct (for a review and references see~\cite{Fields:2006ga}). This number is remarkably similar to the observed amplitude of the dimensionless power spectrum of curvature fluctuations, which leads one to look for a common origin. 

In this Letter we explore the possibility that a net lepton number, sufficient to explain the matter-antimatter asymmetry, may be generated via the production of left-handed standard model neutrinos during inflation. This scenario is naturally accommodated in axion-inflation scenarios, involving a dimension-five derivative coupling of the axion to by-product neutrinos. In our scenario, we assume neutrino masses are generated via the seesaw mechanism, whereby  the standard model neutrinos couple  to GUT-scale  right-handed Majorana neutrinos.  During inflation, this seesaw may also be active, since the electroweak symmetry can be broken by quantum vacuum fluctuations during inflation \cite{ Enqvist:2013kaa}. These quantum fluctuations can generate a  Hubble-scale  {root-mean-square (rms) value} for the Higgs field which leads to the generation of masses for the standard model fields, including neutrinos. The axion-inflaton couples derivatively to the neutrino fields, which leads to the asymmetric production of  neutrino helicity states, while the seesaw mechanism ensures that production of right-handed neutrinos is highly suppressed compared to the production of left-handed neutrinos.  Following inflation, the Higgs  {condensate} decays, the neutrinos become massless, and the resulting helicity asymmetry becomes equivalent to a net lepton number. The result is the net production of lepton number in the form of left-handed by-product neutrinos. The electroweak sphaleron will act on this net lepton number $L$ conserving $B-L$, where $B$ is baryon number, while driving $B+L$ to zero and thus generating a net baryon asymmetry.

The use of rolling scalars coupled to fermionic currents in models for baryogenesis has a long history going back to original work by Cohen and Kaplan \cite{Cohen:1987vi, Cohen:1988kt}.  Dolgov and Freese \cite{Dolgov:1994zq} proposed a rolling axion coupled to the $B-L$ current to generate a baryon asymmetry in the presence of baryon number violating processes during reheating. In contrast to this, our model does not require a complex axion that carries baryon number. As we will explain, in our model the baryon asymmetry is produced via leptogenesis which originates from a helicity asymmetry produced during inflation.
Related ideas have appeared recently such as Higgs relaxation leptogenesis \cite{Kusenko:2014lra, Pearce:2015nga} and axion-oscillation leptogenesis \cite{Kusenko:2014uta}.

We  work in natural units $c = \hbar = k_B =  1$ and denote the reduced Planck mass by $M_{\rm Pl}^{-1} = \sqrt{(8\pi G)}$. We use the 2-component spinor conventions reviewed in \cite{Dreiner:2008tw}, and work with a Friedmann-Robertson-Walker metric with ``mostly plus'' convention.

\section{Axion Inflation with Majorana Fermions}

In addition to the usual action for the standard model of particle physics, we consider a model of pseudoscalar inflation together with a set of Majorana fermions
\begin{align}\label{eqn:axionact}\nonumber
\mathcal{S} = & \int {\rm d}^4 x \Big\{\sqrt{-g}\Big[\frac{M_{\rm Pl}^2}{2}R - \frac{1}{2}(\partial\phi)^2 -  V(\phi)\Big]\\\nn
&+ i\f_i^\dagger{}_{\dot\alpha} e^{\mu}{}_{a}\bar{\sigma}^{a \dot\alpha \beta}\partial_{\mu} \f_{i}{}_{\beta}-\frac{1}{2}m_{ij} (\f_i^{\alpha} \f_j{}_{\alpha} +\f_i^\dagger{}_{\dot\alpha} \f^{\dagger}_{j}{}^{\dot\alpha})\\
 & +\frac{C}{f}\partial_{\mu}\phi \f_i^\dagger{}_{\dot\alpha} e^{\mu}{}_{a}\bar{\sigma}^{a \dot\alpha \beta}\f_{i}{}_{\beta}\Big\}.
\end{align}
Here $\phi$ is the  {real} pseudoscalar (axion) inflaton with a potential $V(\phi)$ which softly breaks the axion shift symmetry and drives a period of slow-roll inflation. The Majorana fermion fields are  $\nu_{i}$, which we have rescaled  by their conformal weight $a^{3/2}$ in order to write the derivative as a partial (rather than covariant) derivative, $e^{\mu}{}_{a}$ are vierbiens \footnote{For a Friedmann-Robertson-Walker metric these are $e^{0}{}_{0} = 1$, $e^i{}_{j} = a^{-1}\delta^{i}{}_j$, with all others zero.}, while $f$ is a mass scale associated with the axion and $C$ is a dimensionless coupling. 

We have written a generic Majorana theory here, but we have in mind the neutrino sector of the standard model augmented with heavy right-handed neutrinos (with Majorana mass terms) to give mass to the by-product neutrinos via the seesaw mechanism. For simplicity, we will later consider only a single generation of neutrinos, and write both the right and left-handed neutrino fields as left-handed spinors; $\nu_{L,i} = \nu_{i}$ ($i\leq 3$). The physical right handed neutrinos are 
$N_{Ri} = \nu^\dagger_{i}$ ($ { 3< i \leq 6  }$). The axial current is conserved for massless fermions, therefore the axion coupling in Eq.\ \eqref{eqn:axionact} has no effect in this limit, contributing only a boundary term to the action.

In local thermal equilibrium, the rolling axion acts as a chemical potential for helicity. While the Universe is not in  thermal equilibrium during inflation, the effect of the coupling of the axion to the neutrinos during these epochs has a similar effect, and biases the gravitational production of one helicity over the other \cite{Adshead:2015kza}.  For Dirac fermions, where the masses are degenerate $m_i = m$, conservation of charge associated with the additional $U(1)$ symmetry means that this particle production results in a helicity asymmetry but not a matter-antimatter asymmetry. In contrast, for Majorana fermions, with lepton-number violating mass terms, this coupling leads to a helicity asymmetry which is equivalent to lepton number and a matter-antimatter asymmetry if these fields subsequently become  nearly massless.

\section{The standard model Higgs field during inflation}

The standard model Higgs, $\Phi$, has a tree-level potential of the form
\begin{align}
V(\Phi) = \mu^2 \Phi^\dagger \Phi + \lambda (\Phi^\dagger \Phi)^2.
\end{align}
At the electroweak scale, the parameters  $\mu$ and $\lambda$ yield a stable minimum at a VEV of $v_{\rm EW} = 246$ GeV. While these parameters are constant at tree level, they are modified by  both  loop  and  finite  temperature  corrections.  The experimentally preferred top quark and Higgs boson masses give  loop  corrections  that result in a negative running of the coupling $\lambda$ at a sufficiently large  vacuum-expectation-value \cite{Degrassi:2012ry}. For the central values of the standard model parameters the electroweak vacuum is metastable, however, for values within the 2-sigma regions, a stable potential can be achieved. 

During inflation, in the absence of new physics that significantly changes the running of $\lambda$ between the electroweak and inflationary scales, this negative running means that the Higgs field is generically light. Consequently, quantum fluctuations of the Higgs field that are produced during the inflationary epoch are sufficient to generate an  {rms} value \cite{Starobinsky:1994bd, Enqvist:2013kaa}. It is then quite natural to assume that the Higgs field generically has a large  {rms value} during the inflationary epoch, whose size is expected to be of the order \cite{Enqvist:2013kaa}
\begin{align}
\langle \Phi \rangle =\frac{1}{\sqrt{2}} \left(\begin{array}{c}h\\0\end{array}\), \quad h\sim 0.36\frac{H}{\lambda_*^{1/4}},
\end{align}
where $\lambda_*$ is the Higgs self-coupling evaluated at the inflationary energy scale. 
The Higgs condensate breaks electroweak symmetry  during inflation, and the standard model fields acquire masses set by the Hubble scale through the Higgs mechanism.

For simplicity, we will consider only a single left-handed, and a single right-handed neutrino and take the form of the neutrino mass matrix to be
\begin{align}
m_{ij} = \(\begin{array}{cc} 0 & m_D\\ m_D & M \end{array}\), \quad m_D = \frac{y h}{\sqrt{2}},
\end{align}
where $y$ is the Yukawa coupling and $M \sim 10^{16}$ GeV is the right-handed neutrino mass.

To proceed, we work in a basis of mass eigenstates, we diagonalize the fermion sector by rotating the fermion fields. After diagonalization, for a mass hierarchy $M \gg  m_D$, the seesaw mechanism results in a mass matrix with masses $m_i \sim M, m_D^2/M$. We assume that the right-handed neutrinos are heavy compared to all scales of interest, and can be safely neglected. In what follows we  consider only the left-handed neutrinos, whose mass we  take to be a free parameter of order the Hubble scale.

\section{Left-handed neutrino production during inflation}

As noted above, for simplicity we focus on a single generation of neutrinos. Varying the action with respect to $\nu^\dagger$ yields the equation of motion for $\nu$
\begin{align}\label{eqn:chieom}
 \(i e^{\mu}{}_{a} \bar{\sigma}^{a \dot{\alpha}\beta}\partial_{\mu}+\frac{C}{f}\partial_{\mu}\phi e^{\mu}{}_{a}  \bar{\sigma}^{a \dot\alpha \beta}\)\nu_{\beta} = m\nu^{\dagger\dot\alpha},
\end{align}
where $m = m_D^2/M$. We expand each field into a Fourier basis,
\begin{align}\label{eqn:chimode}
\nu_{\alpha}
 = \sum_\lambda \int \frac{{\rm d}^3 k}{(2\pi)^3}\[ x^\lambda _{\alpha}({\bf k}, t)a_{\bf k}^\lambda  e^{i{\bf k}\cdot{\bf x}}+ y^\lambda _{\alpha}({\bf k}, t)a^\dagger{}^\lambda _{\bf k} e^{-i{\bf k}\cdot{\bf x}} \],
\end{align}
where we have introduced  creation and annihilation operators, $a_{ \bf k}^\lambda$ and $a_{ \bf k}^{\dagger\lambda}$, which  satisfy the  anticommutation relations
$\{a^\lambda _{\bf k}, a^{\dagger \lambda '}_{\bf k'}\} = (2\pi)^3\delta^3({\bf k}- {\bf k}')\delta_{\lambda \lambda '}$,
with all other anticommutators vanishing, as usual.  We quantize the fields by imposing the anticommutation relations
$\{\nu_{\alpha}({\bf x}, t), \pi^\beta_{\nu} ({\bf y}, t) \} = i\delta^{\beta}_{\alpha} \delta^{3}({\bf x} - {\bf y})$.
The canonical momenta of the fermions are found in the usual way
\begin{align}
\pi^\beta_{\nu}  = \frac{\partial\mathcal{L}}{\partial \dot\nu_{ \beta}}= & i \nu^\dagger{}_{\dot\alpha} \bar{\sigma}^{0\dot\alpha \beta},
\end{align}
where an overdot here and  {throughout} denotes a derivative with respect to cosmic time, $t$.
We work in a basis of helicity eigenspinors, which satisfy
\begin{align}
\vec{\sigma}\cdot\hat{k} \xi_\lambda = \lambda \xi_\lambda, \quad \lambda = \pm 1, \quad \xi_{-\lambda}(-\hat{k}) = & \iota^{\lambda}_{\hat{k}}\xi_{\lambda}(\hat{k}),
\end{align}
where $\iota^{\lambda}_{\hat{k}}$ is a phase that satisfies
$
\iota^{\lambda *}_{\hat{k}}\iota^{\lambda}_{\hat{k}} = 1$, and  $\iota^{\lambda}_{-\hat{k}} = -\iota^{\lambda}_{\hat{k}}$. 
Writing the spinors in this helicity basis as
\begin{align}\label{eqn:helicity}
x_{ \alpha}^{\lambda}({\bf k}, t) = & X^{\lambda}_{k} (t) \xi_\lambda({\bf k}), \quad
y^{\lambda \dagger \dot\alpha }({\bf k}, t) =  Y^{\lambda *}_{ k}(t) \xi_\lambda({\bf k}),
\end{align}
canonical quantization requires that the fermion wave functions satisfy
\begin{align}
\sum_\lambda\[ X^{\lambda}_{k} (t) X^{\lambda *}_{k} (t) + Y^{\lambda *}_{ k}(t) Y^{\lambda}_{k}(t)  \] = 1.
\end{align}
Substituting Eqs.\ \eqref{eqn:helicity} and \eqref{eqn:chimode} into Eq.\ \eqref{eqn:chieom}, the equations of motion for the fermion wave functions are
 \begin{align}\nn\label{eqn:eomsmaj1st}
i \left [   \partial_{t}- i \(\frac{k}{a} \lambda+\frac{C}{f}\dot\phi\) \right ] X^{\lambda}_{ k} (t)  = & m Y^{\lambda *}_{k}(t) ,
\\
 i \left [ \partial_{t} + i \(\frac{k}{a} \lambda+\frac{C}{f}\dot\phi\)  \right ] Y^{\lambda*}_{k}(t)= & m X^{\lambda}_{k} (t).
\end{align}
Note that the effect of the axion coupling is a helicity-dependent shift in the effective wavenumber of the modes 
$
k/a \lambda \to k/a \lambda+C\dot\phi/f.
$

Assuming approximate de Sitter space during inflation, $H\approx$ const., and taking $\dot{\phi}/H \approx$ const., the canonically normalized solutions $X_k^\lambda(t), Y_k^{\lambda * }(t)$ of Eq.\ \eqref{eqn:eomsmaj1st} that match onto the Bunch-Davies vacuum \cite{Adshead:2015kza} are
\begin{align}\nn \label{eqn:modessols}
X^\lambda_{ k}(k\tau) =  & \( - \frac{i m}{H}\)^{\frac{\lambda+1}{2}}  \frac{e^{i\theta}e^{-\lambda\frac{\pi}{2}\vartheta}}{\sqrt{2k\tau}}W_{-\lambda\(\frac{1}{2}+i\vartheta\),\mu}(2 i k\tau),\\\nn
 Y^{\lambda*}_{k}(k\tau) = & \( - \frac{i m}{H}\)^{\frac{-\lambda+1}{2}} \frac{e^{i\theta'}e^{-\lambda\frac{\pi}{2}\vartheta}}{\sqrt{2k\tau}} W_{\lambda\(\frac{1}{2}-i\vartheta\), \mu}(2 i  k\tau) , 
 \end{align}
where  $W_{\kappa, \mu}(x)$ are the Whittaker W functions, $\tau = -H^{-1}e^{Ht}$, $\theta$ and $\theta'$ are arbitrary phases, and 
 \begin{align}
 \vartheta =-  \frac{C}{f}\frac{\dot\phi}{H},& \quad \mu = \sqrt{\frac{m^2}{H^2} + \vartheta^2}.
\end{align}
In order to determine the particle number at any given time, we perform a Bogoliubov transformation, and compare the exact wave functions to an instantaneous WKB solution that diagonalizes the Hamiltonian. This instantaneous quasiparticle number is \cite{Greene:2000ew, Peloso:2000hy}
\begin{align}
n_{\nu}^\lambda(k) =\frac{\[|\dot{X}^\lambda_{k}|^2 + \omega_{\lambda}^2|{X}^\lambda_{k}|^2 - 2 \omega_\lambda \Im (X^\lambda_{k}\dot{X}^{\lambda*}_{k})\]}{\omega_\lambda(\tilde{k}_\lambda + \omega_\lambda)}.
\end{align}
where the effective frequency and wavenumber are
\begin{align}\label{eqn:fermionfreq}
\omega^2_\lambda(t) = &  \tilde{k}_{\lambda}(t)^2 + m^2,\quad \tilde{k}_{\lambda}(t) =  \(\frac{k}{a}\lambda +\frac{C}{f}\dot\phi \).
\end{align}
Using the analytic expressions for the wave functions  from Eq.\ \eqref{eqn:modessols}, we derive analytic expressions for the quasiparticle number during inflation.
Assuming that $m \neq 0$ and taking the limit $k/aH \to 0$, we find
\begin{align}\label{inflationnumber}
n_{\nu}^{\pm}(k) = &e^{-\pi\(\mp\vartheta + \sqrt{\frac{m^2}{H^2}+\vartheta^2}\)}\frac{\sinh\[\pi\( \sqrt{\frac{m^2}{H^2}+\vartheta^2}\pm\vartheta\)\]}{\sinh\[2\pi\( \sqrt{\frac{m^2}{H^2}+\vartheta^2}\)\]}.
\end{align}
 {While Eq.\ \eqref{inflationnumber} is derived in the limit $k/aH \to 0$, for $m \ll \vartheta H$ it is a good approximation for the particle number for modes that satisfy $k/aH <\vartheta$. }
Note that in the absence of the coupling to the axion ($\vartheta = 0$), production of both helicity states is symmetric, as expected, and highly suppressed for fermions with masses larger than the Hubble rate. For small masses, the occupation number approaches its maximum value of $1/2$ as $m/H \to 0$. However, for $m= 0$, the theory is conformally equivalent to a Minkowski-space theory, and no particle production occurs \footnote{Eq. \eqref{inflationnumber} has been derived by assuming $m / H \neq 0$, and taking the limit $k/(aH) \to 0$. At $m = 0$, the Hamiltonian is always diagonal, and the Bogoliubov transformation is trivial.}.  When the coupling to the axion is switched on, the particle production is asymmetric between the helicity states. For $\vartheta > 0$ ($\vartheta < 0$), particle production of the $\lambda = +$ ($\lambda = -$) helicity state is enhanced while particle production of the $\lambda = -$ ($\lambda = +$) mode is suppressed. Larger couplings allow the production of fermions with increasingly large mass.

 {This particle production can be understood by studying the evolution of Eq.\ \eqref{eqn:eomsmaj1st} in the WKB approximation. For $m/H \ll \vartheta$, particle production occurs near the points of nonadiabatic evolution of $\tilde k$. As discussed in \cite{Adshead:2015kza},  these occur whenever $\tilde{k} = 0$, where $k\lambda/a=-(C/f) \dot \phi$. After this event, $k/aH < \vartheta$, and the resulting (quasi-)particle number is approximately constant. 
The maximum comoving wavenumber of excited fermion states can be calculated as the maximum comoving wavenumber that makes $\tilde k=0$. By simple inspection of the terms in $\tilde k$, we can see the scaling $k_{\rm max}\propto (C/f)\dot \phi$. The proportionality factor depends on the axion potential, as explained in \cite{Adshead:2015kza}.}

At the end of inflation the resulting helicity asymmetry is (assuming $\vartheta >0$)
\begin{align} \label{eqn:helicityasym}
n^{h}_{\nu} &= \sum_{\lambda = \pm 1}\frac{3\lambda }{2\pi^2a^3}  \int_0^{\infty}\!\!\!\!\! n_\nu^\lambda(k) k^2 {\rm d} k 
\approx \frac{\langle n^{+}_\nu \rangle}{2\pi^2 a^3} \(\frac{C\phi_0 }{f}a_{\rm e}H_{\rm e}\)^3, 
\end{align}
where we have used $k_{\rm max} \propto (C\phi_0/f)a_e H_e$, and $a_{\rm e}$ and $H_{\rm e}$ are the scale factor and Hubble rate, respectively, at the end of inflation. The quantity $\langle n^{\lambda}_\nu \rangle$ is the phase-space-averaged occupation number  \cite{Adshead:2015kza} of the helicity $\lambda$. We have taken $\dot\phi/H = \phi_0$, where $\phi_0 \sim M_{\rm pl}$ is the field value at which inflation ends and oscillations begin. We have also summed over the three generations of neutrinos. 

After inflation, once the Hubble rate drops below the mass of the Higgs field, the Higgs   {condensate} will decay restoring electroweak symmetry  \cite{Enqvist:2013kaa} and this helicity asymmetry will become equal to lepton number. 

Following inflation the axion oscillates which results in the production of both helicity states of the left-handed neutrinos. However, the helicity states that are produced during inflation (and during  {the} first axion zero crossing) are produced out to a larger wavenumber due to the fact that the other helicity is not produced until the axion velocity changes sign. Provided that the Higgs   {condensate} does not decay immediately following inflation, the subsequent production events are less efficient due to Hubble damping of the axion velocity. The end result is that, even after the axion oscillations are taken into account, a helicity asymmetry of the order of  Eq.\ \eqref{eqn:helicityasym} is generic for a wide range of parameter space.  A detailed analysis of fermion production during and after axion inflation appears in \cite{Adshead:2015kza}.

\section{Baryon-to-photon ratio}

During reheating, the energy density $\rho = 3M_{\rm Pl}^2 H_e^2$ in the inflaton is converted into radiation, over an epoch where the scale factor expands from $a_{\rm e}$ at the end of inflation to $a_{\rm R}$ at the end of reheating.  
 {Taking an arbitrary equation of state ($w$) of the Universe between the end of inflation and reheating,}
we write the comoving entropy at reheating as
\begin{equation}\label{eqn:comovingentropy}
a_{\rm R}^3 s =  \frac{2\pi^2}{45}g_* a_{\rm R}^3 T_R^3  = \frac{    {4}   M_{\rm Pl}^2H_e^2   {a_e^3  }}{T_R}   {  \left ( {a_e\over a_R}   \right )^{3w} },
\end{equation}
where $T_{\rm R} \sim \sqrt{\Gamma_I M_{\rm Pl}}$ is the reheat temperature, $\Gamma_I$ the inflaton decay rate, and $g_*$ is the effective number of relativistic degrees of freedom. In obtaining this expression,  we used the Friedmann equation to relate the energy density at the end of inflation to the energy density at reheating
\begin{align}
3 M_{\rm Pl}^2 H_{\rm R}^2  = \rho_{e}   { \left ( \frac{a_{\rm e}}{a_{\rm R}} \right )^{3(1+w)} }= \frac{\pi^2}{30}g_* T_{R}^4,
\end{align}
where $H_{\rm R}$ is the Hubble rate at reheating, and $\rho_e = 3 M_{\rm Pl}^2 H_e^2$ is the energy density at the end of inflation. Using Eq.\ \eqref{eqn:comovingentropy}, we arrive at an estimate for the asymmetry parameter at reheating
\begin{align}
\eta_{\rm R}\approx  \frac{\langle n_\nu \rangle}{ {8}\pi^2} \(\frac{C\phi_0 }{f}\)^3\frac{H_{\rm e}}{M_{\rm Pl}}\frac{T_{\rm R}}{M_{\rm Pl}}  {  \left ( {a_R\over a_e}   \right )^{3w} }  .
\end{align}
Following reheating, by-product sphaleron processes redistribute this asymmetry between lepton and baryon numbers \cite{Kuzmin:1985mm}. Standard model entropy generation and the redistribution of the lepton number into the baryon number lowers $\eta_{\rm R}$ by $1$ or $2$ orders of magnitude, which implies we require $\eta_{\rm R} \sim 10^{-7} - 10^{-8}$ in order to explain the present day baryon asymmetry.

In order to obtain the correct asymmetry, assuming that the average occupation number $\langle n^{+}_{\nu} \rangle$ is of order unity, and that the Universe is matter dominated during reheating $w = 0$, we see that we require both high-scale inflation and a relatively high reheating temperature in order to overcome the $M_{\rm Pl}^{-2}$ suppression. However, the tensor-to-scalar ratio constrains the inflationary energy scale to be $H \lesssim 10^{-5} - 10^{-6}$ $M_{\rm Pl}$. On the other hand, in order to prevent the washout of the lepton number stored in the left-handed neutrinos, we need to ensure that reheating occurs below the scale at which lepton-number violating processes involving the exchange of heavy right-handed neutrinos are in equilibrium. In order to prevent excessive washout, the reheat temperature must be  below $T_R \lesssim 3 \times 10^{14}$ GeV  \cite{Kusenko:2014lra}. 

Taking $T_{R} \sim 10^{13}$ GeV, GUT scale inflation $H_e \sim 10^{-6}$ $M_{\rm Pl}$, we see that for $f \sim 10^{-2}-10^{-3}$ $M_{\rm Pl}$ we require order unity couplings $C \sim 1$  between the axion and by-product neutrinos to generate a baryon asymmetry of the appropriate size today.

\section{Discussion and Conclusions}

In this Letter we have studied the generation of lepton number via the biased production of left-handed neutrino helicity states during inflation. In our scenario, we have assumed that quantum effects break electroweak symmetry during inflation and the by-product neutrinos are given masses by GUT-scale right-handed Majorana neutrinos. Following inflation, electroweak symmetry is restored as the Higgs  {condensate} decays and the left-handed neutrino helicity becomes equivalent to a net lepton number. Sphaleron processes redistribute this net lepton number, $L$, into baryon number $B$,  via processes that violate $B+L$ but conserve $B-L$ resulting in a net baryon number.  

The number density of neutrinos produced by this mechanism is proportional to the cube of the Hubble rate at the end of inflation. This is simply due to the nature of inflationary particle production, which populates states with momenta near the Hubble scale. High-scale inflation at or near the GUT scale produces the smallest allowable Hubble length and, therefore the largest number density of neutrinos.   In order that these neutrinos are not diluted too much by the subsequent expansion between the end of inflation and the onset of the hot big-bang phase, the reheat temperature needs to be high. However, this reheating temperature cannot be significantly higher than $\sim 10^{14}$ GeV, as scattering processes mediated by heavy right-handed neutrinos can wash out the asymmetry. 

To produce the observed baryon asymmetry with this mechanism, we require the Hubble rate at the end of inflation $H_{e} \sim 10^{-6}M_{\rm Pl}$, a reheat temperature $T_{R} \sim 10^{13}$ GeV and a derivative coupling between an axionic inflaton and a Majorana neutrino sector with strength $C/f \sim 10^{2}-10^3 M_{\rm Pl}^{-1}$.

A number of details of this scenario still need to be explored. First, since the fermions are produced during inflation, their backreaction could lead to significant contributions to the curvature spectrum. Second, we have made only rough estimates of how much of the helicity asymmetry eventually becomes a final baryon asymmetry.  This process is dependent on the details of reheating, and requires modeling the decay of the Higgs   {condensate} and solving Boltzmann equations. 
 {Finally, the possibility of fermion isocurvature perturbations \cite{Chung:2013rda} could lead to an independent observable that is correlated to the baryon asymmetry.}
We leave these studies for future work.

{\bf Acknowledgements: } We thank Jessie Shelton, Emil Martinec, and Cliff Burgess for useful conversations.  E.I.S. acknowledges support from a Fortner fellowship.  P.A. acknowledges support by National Science Foundation Grant No. PHYS-1066293 and the hospitality of the Aspen Center for Physics.

\appendix

\bibliography{Lepto}

\begin{thebibliography}{23}%
\makeatletter
\providecommand \@ifxundefined [1]{%
 \@ifx{#1\undefined}
}%
\providecommand \@ifnum [1]{%
 \ifnum #1\expandafter \@firstoftwo
 \else \expandafter \@secondoftwo
 \fi
}%
\providecommand \@ifx [1]{%
 \ifx #1\expandafter \@firstoftwo
 \else \expandafter \@secondoftwo
 \fi
}%
\providecommand \natexlab [1]{#1}%
\providecommand \enquote  [1]{``#1''}%
\providecommand \bibnamefont  [1]{#1}%
\providecommand \bibfnamefont [1]{#1}%
\providecommand \citenamefont [1]{#1}%
\providecommand \href@noop [0]{\@secondoftwo}%
\providecommand \href [0]{\begingroup \@sanitize@url \@href}%
\providecommand \@href[1]{\@@startlink{#1}\@@href}%
\providecommand \@@href[1]{\endgroup#1\@@endlink}%
\providecommand \@sanitize@url [0]{\catcode `\\12\catcode `\$12\catcode
  `\&12\catcode `\#12\catcode `\^12\catcode `\_12\catcode `\%12\relax}%
\providecommand \@@startlink[1]{}%
\providecommand \@@endlink[0]{}%
\providecommand \url  [0]{\begingroup\@sanitize@url \@url }%
\providecommand \@url [1]{\endgroup\@href {#1}{\urlprefix }}%
\providecommand \urlprefix  [0]{URL }%
\providecommand \Eprint [0]{\href }%
\providecommand \doibase [0]{http://dx.doi.org/}%
\providecommand \selectlanguage [0]{\@gobble}%
\providecommand \bibinfo  [0]{\@secondoftwo}%
\providecommand \bibfield  [0]{\@secondoftwo}%
\providecommand \translation [1]{[#1]}%
\providecommand \BibitemOpen [0]{}%
\providecommand \bibitemStop [0]{}%
\providecommand \bibitemNoStop [0]{.\EOS\space}%
\providecommand \EOS [0]{\spacefactor3000\relax}%
\providecommand \BibitemShut  [1]{\csname bibitem#1\endcsname}%
\let\auto@bib@innerbib\@empty
\bibitem [{\citenamefont {Freese}\ \emph {et~al.}(1990)\citenamefont {Freese},
  \citenamefont {Frieman},\ and\ \citenamefont {Olinto}}]{Freese:1990rb}%
  \BibitemOpen
  \bibfield  {author} {\bibinfo {author} {\bibfnamefont {K.}~\bibnamefont
  {Freese}}, \bibinfo {author} {\bibfnamefont {J.~A.}\ \bibnamefont {Frieman}},
  \ and\ \bibinfo {author} {\bibfnamefont {A.~V.}\ \bibnamefont {Olinto}},\
  }\href {\doibase 10.1103/PhysRevLett.65.3233} {\bibfield  {journal} {\bibinfo
   {journal} {Phys.Rev.Lett.}\ }\textbf {\bibinfo {volume} {65}},\ \bibinfo
  {pages} {3233} (\bibinfo {year} {1990})}\BibitemShut {NoStop}%
\bibitem [{\citenamefont {Silverstein}\ and\ \citenamefont
  {Westphal}(2008)}]{Silverstein:2008sg}%
  \BibitemOpen
  \bibfield  {author} {\bibinfo {author} {\bibfnamefont {E.}~\bibnamefont
  {Silverstein}}\ and\ \bibinfo {author} {\bibfnamefont {A.}~\bibnamefont
  {Westphal}},\ }\href {\doibase 10.1103/PhysRevD.78.106003} {\bibfield
  {journal} {\bibinfo  {journal} {Phys.Rev.}\ }\textbf {\bibinfo {volume}
  {D78}},\ \bibinfo {pages} {106003} (\bibinfo {year} {2008})},\ \Eprint
  {http://arxiv.org/abs/0803.3085} {arXiv:0803.3085 [hep-th]} \BibitemShut
  {NoStop}%
\bibitem [{\citenamefont {Ade}\ \emph {et~al.}(2014{\natexlab{a}})\citenamefont
  {Ade} \emph {et~al.}}]{Ade:2014xna}%
  \BibitemOpen
  \bibfield  {author} {\bibinfo {author} {\bibfnamefont {P.~A.~R.}\
  \bibnamefont {Ade}} \emph {et~al.} (\bibinfo {collaboration} {BICEP2}),\
  }\href {\doibase 10.1103/PhysRevLett.112.241101} {\bibfield  {journal}
  {\bibinfo  {journal} {Phys. Rev. Lett.}\ }\textbf {\bibinfo {volume} {112}},\
  \bibinfo {pages} {241101} (\bibinfo {year} {2014}{\natexlab{a}})},\ \Eprint
  {http://arxiv.org/abs/1403.3985} {arXiv:1403.3985 [astro-ph.CO]} \BibitemShut
  {NoStop}%
\bibitem [{\citenamefont {Ade}\ \emph {et~al.}(2014{\natexlab{b}})\citenamefont
  {Ade} \emph {et~al.}}]{Ade:2013zuv}%
  \BibitemOpen
  \bibfield  {author} {\bibinfo {author} {\bibfnamefont {P.~A.~R.}\
  \bibnamefont {Ade}} \emph {et~al.} (\bibinfo {collaboration} {Planck}),\
  }\href {\doibase 10.1051/0004-6361/201321591} {\bibfield  {journal} {\bibinfo
   {journal} {Astron. Astrophys.}\ }\textbf {\bibinfo {volume} {571}},\
  \bibinfo {pages} {A16} (\bibinfo {year} {2014}{\natexlab{b}})},\ \Eprint
  {http://arxiv.org/abs/1303.5076} {arXiv:1303.5076 [astro-ph.CO]} \BibitemShut
  {NoStop}%
\bibitem [{\citenamefont {Komatsu}\ \emph {et~al.}(2011)\citenamefont {Komatsu}
  \emph {et~al.}}]{Komatsu:2010fb}%
  \BibitemOpen
  \bibfield  {author} {\bibinfo {author} {\bibfnamefont {E.}~\bibnamefont
  {Komatsu}} \emph {et~al.} (\bibinfo {collaboration} {WMAP Collaboration}),\
  }\href {\doibase 10.1088/0067-0049/192/2/18} {\bibfield  {journal} {\bibinfo
  {journal} {Astrophys.J.Suppl.}\ }\textbf {\bibinfo {volume} {192}},\ \bibinfo
  {pages} {18} (\bibinfo {year} {2011})},\ \Eprint
  {http://arxiv.org/abs/1001.4538} {arXiv:1001.4538 [astro-ph.CO]} \BibitemShut
  {NoStop}%
\bibitem [{\citenamefont {Fields}\ and\ \citenamefont
  {Sarkar}(2006)}]{Fields:2006ga}%
  \BibitemOpen
  \bibfield  {author} {\bibinfo {author} {\bibfnamefont {B.}~\bibnamefont
  {Fields}}\ and\ \bibinfo {author} {\bibfnamefont {S.}~\bibnamefont
  {Sarkar}},\ }\href@noop {} {\  (\bibinfo {year} {2006})},\ \Eprint
  {http://arxiv.org/abs/astro-ph/0601514} {arXiv:astro-ph/0601514 [astro-ph]}
  \BibitemShut {NoStop}%
\bibitem [{\citenamefont {Enqvist}\ \emph {et~al.}(2013)\citenamefont
  {Enqvist}, \citenamefont {Meriniemi},\ and\ \citenamefont
  {Nurmi}}]{Enqvist:2013kaa}%
  \BibitemOpen
  \bibfield  {author} {\bibinfo {author} {\bibfnamefont {K.}~\bibnamefont
  {Enqvist}}, \bibinfo {author} {\bibfnamefont {T.}~\bibnamefont {Meriniemi}},
  \ and\ \bibinfo {author} {\bibfnamefont {S.}~\bibnamefont {Nurmi}},\ }\href
  {\doibase 10.1088/1475-7516/2013/10/057} {\bibfield  {journal} {\bibinfo
  {journal} {JCAP}\ }\textbf {\bibinfo {volume} {1310}},\ \bibinfo {pages}
  {057} (\bibinfo {year} {2013})},\ \Eprint {http://arxiv.org/abs/1306.4511}
  {arXiv:1306.4511 [hep-ph]} \BibitemShut {NoStop}%
\bibitem [{\citenamefont {Cohen}\ and\ \citenamefont
  {Kaplan}(1987)}]{Cohen:1987vi}%
  \BibitemOpen
  \bibfield  {author} {\bibinfo {author} {\bibfnamefont {A.~G.}\ \bibnamefont
  {Cohen}}\ and\ \bibinfo {author} {\bibfnamefont {D.~B.}\ \bibnamefont
  {Kaplan}},\ }\href {\doibase 10.1016/0370-2693(87)91369-4} {\bibfield
  {journal} {\bibinfo  {journal} {Phys.Lett.}\ }\textbf {\bibinfo {volume}
  {B199}},\ \bibinfo {pages} {251} (\bibinfo {year} {1987})}\BibitemShut
  {NoStop}%
\bibitem [{\citenamefont {Cohen}\ and\ \citenamefont
  {Kaplan}(1988)}]{Cohen:1988kt}%
  \BibitemOpen
  \bibfield  {author} {\bibinfo {author} {\bibfnamefont {A.~G.}\ \bibnamefont
  {Cohen}}\ and\ \bibinfo {author} {\bibfnamefont {D.~B.}\ \bibnamefont
  {Kaplan}},\ }\href {\doibase 10.1016/0550-3213(88)90134-4} {\bibfield
  {journal} {\bibinfo  {journal} {Nucl.Phys.}\ }\textbf {\bibinfo {volume}
  {B308}},\ \bibinfo {pages} {913} (\bibinfo {year} {1988})}\BibitemShut
  {NoStop}%
\bibitem [{\citenamefont {Dolgov}\ and\ \citenamefont
  {Freese}(1995)}]{Dolgov:1994zq}%
  \BibitemOpen
  \bibfield  {author} {\bibinfo {author} {\bibfnamefont {A.}~\bibnamefont
  {Dolgov}}\ and\ \bibinfo {author} {\bibfnamefont {K.}~\bibnamefont
  {Freese}},\ }\href {\doibase 10.1103/PhysRevD.51.2693} {\bibfield  {journal}
  {\bibinfo  {journal} {Phys.Rev.}\ }\textbf {\bibinfo {volume} {D51}},\
  \bibinfo {pages} {2693} (\bibinfo {year} {1995})},\ \Eprint
  {http://arxiv.org/abs/hep-ph/9410346} {arXiv:hep-ph/9410346 [hep-ph]}
  \BibitemShut {NoStop}%
\bibitem [{\citenamefont {Kusenko}\ \emph
  {et~al.}(2015{\natexlab{a}})\citenamefont {Kusenko}, \citenamefont {Pearce},\
  and\ \citenamefont {Yang}}]{Kusenko:2014lra}%
  \BibitemOpen
  \bibfield  {author} {\bibinfo {author} {\bibfnamefont {A.}~\bibnamefont
  {Kusenko}}, \bibinfo {author} {\bibfnamefont {L.}~\bibnamefont {Pearce}}, \
  and\ \bibinfo {author} {\bibfnamefont {L.}~\bibnamefont {Yang}},\ }\href
  {\doibase 10.1103/PhysRevLett.114.061302} {\bibfield  {journal} {\bibinfo
  {journal} {Phys.Rev.Lett.}\ }\textbf {\bibinfo {volume} {114}},\ \bibinfo
  {pages} {061302} (\bibinfo {year} {2015}{\natexlab{a}})},\ \Eprint
  {http://arxiv.org/abs/1410.0722} {arXiv:1410.0722 [hep-ph]} \BibitemShut
  {NoStop}%
\bibitem [{\citenamefont {Pearce}\ \emph {et~al.}(2015)\citenamefont {Pearce},
  \citenamefont {Yang}, \citenamefont {Kusenko},\ and\ \citenamefont
  {Peloso}}]{Pearce:2015nga}%
  \BibitemOpen
  \bibfield  {author} {\bibinfo {author} {\bibfnamefont {L.}~\bibnamefont
  {Pearce}}, \bibinfo {author} {\bibfnamefont {L.}~\bibnamefont {Yang}},
  \bibinfo {author} {\bibfnamefont {A.}~\bibnamefont {Kusenko}}, \ and\
  \bibinfo {author} {\bibfnamefont {M.}~\bibnamefont {Peloso}},\ }\href
  {\doibase 10.1103/PhysRevD.92.023509} {\bibfield  {journal} {\bibinfo
  {journal} {Phys. Rev.}\ }\textbf {\bibinfo {volume} {D92}},\ \bibinfo {pages}
  {023509} (\bibinfo {year} {2015})},\ \Eprint
  {http://arxiv.org/abs/1505.02461} {arXiv:1505.02461 [hep-ph]} \BibitemShut
  {NoStop}%
\bibitem [{\citenamefont {Kusenko}\ \emph
  {et~al.}(2015{\natexlab{b}})\citenamefont {Kusenko}, \citenamefont
  {Schmitz},\ and\ \citenamefont {Yanagida}}]{Kusenko:2014uta}%
  \BibitemOpen
  \bibfield  {author} {\bibinfo {author} {\bibfnamefont {A.}~\bibnamefont
  {Kusenko}}, \bibinfo {author} {\bibfnamefont {K.}~\bibnamefont {Schmitz}}, \
  and\ \bibinfo {author} {\bibfnamefont {T.~T.}\ \bibnamefont {Yanagida}},\
  }\href {\doibase 10.1103/PhysRevLett.115.011302} {\bibfield  {journal}
  {\bibinfo  {journal} {Phys. Rev. Lett.}\ }\textbf {\bibinfo {volume} {115}},\
  \bibinfo {pages} {011302} (\bibinfo {year} {2015}{\natexlab{b}})},\ \Eprint
  {http://arxiv.org/abs/1412.2043} {arXiv:1412.2043 [hep-ph]} \BibitemShut
  {NoStop}%
\bibitem [{\citenamefont {Dreiner}\ \emph {et~al.}(2010)\citenamefont
  {Dreiner}, \citenamefont {Haber},\ and\ \citenamefont
  {Martin}}]{Dreiner:2008tw}%
  \BibitemOpen
  \bibfield  {author} {\bibinfo {author} {\bibfnamefont {H.~K.}\ \bibnamefont
  {Dreiner}}, \bibinfo {author} {\bibfnamefont {H.~E.}\ \bibnamefont {Haber}},
  \ and\ \bibinfo {author} {\bibfnamefont {S.~P.}\ \bibnamefont {Martin}},\
  }\href {\doibase 10.1016/j.physrep.2010.05.002} {\bibfield  {journal}
  {\bibinfo  {journal} {Phys.Rept.}\ }\textbf {\bibinfo {volume} {494}},\
  \bibinfo {pages} {1} (\bibinfo {year} {2010})},\ \Eprint
  {http://arxiv.org/abs/0812.1594} {arXiv:0812.1594 [hep-ph]} \BibitemShut
  {NoStop}%
\bibitem [{Note1()}]{Note1}%
  \BibitemOpen
  \bibinfo {note} {For a Friedmann-Robertson-Walker metric these are
  $e^{0}{}_{0} = 1$, $e^i{}_{j} = a^{-1}\delta ^{i}{}_j$, with all others
  zero.}\BibitemShut {Stop}%
\bibitem [{\citenamefont {Adshead}\ and\ \citenamefont
  {Sfakianakis}(2015)}]{Adshead:2015kza}%
  \BibitemOpen
  \bibfield  {author} {\bibinfo {author} {\bibfnamefont {P.}~\bibnamefont
  {Adshead}}\ and\ \bibinfo {author} {\bibfnamefont {E.~I.}\ \bibnamefont
  {Sfakianakis}},\ }\href {\doibase 10.1088/1475-7516/2015/11/021} {\bibfield
  {journal} {\bibinfo  {journal} {JCAP}\ }\textbf {\bibinfo {volume} {1511}},\
  \bibinfo {pages} {021} (\bibinfo {year} {2015})},\ \Eprint
  {http://arxiv.org/abs/1508.00891} {arXiv:1508.00891 [hep-ph]} \BibitemShut
  {NoStop}%
\bibitem [{\citenamefont {Degrassi}\ \emph {et~al.}(2012)\citenamefont
  {Degrassi}, \citenamefont {Di~Vita}, \citenamefont {Elias-Miro},
  \citenamefont {Espinosa}, \citenamefont {Giudice} \emph
  {et~al.}}]{Degrassi:2012ry}%
  \BibitemOpen
  \bibfield  {author} {\bibinfo {author} {\bibfnamefont {G.}~\bibnamefont
  {Degrassi}}, \bibinfo {author} {\bibfnamefont {S.}~\bibnamefont {Di~Vita}},
  \bibinfo {author} {\bibfnamefont {J.}~\bibnamefont {Elias-Miro}}, \bibinfo
  {author} {\bibfnamefont {J.~R.}\ \bibnamefont {Espinosa}}, \bibinfo {author}
  {\bibfnamefont {G.~F.}\ \bibnamefont {Giudice}},  \emph {et~al.},\ }\href
  {\doibase 10.1007/JHEP08(2012)098} {\bibfield  {journal} {\bibinfo  {journal}
  {JHEP}\ }\textbf {\bibinfo {volume} {1208}},\ \bibinfo {pages} {098}
  (\bibinfo {year} {2012})},\ \Eprint {http://arxiv.org/abs/1205.6497}
  {arXiv:1205.6497 [hep-ph]} \BibitemShut {NoStop}%
\bibitem [{\citenamefont {Starobinsky}\ and\ \citenamefont
  {Yokoyama}(1994)}]{Starobinsky:1994bd}%
  \BibitemOpen
  \bibfield  {author} {\bibinfo {author} {\bibfnamefont {A.~A.}\ \bibnamefont
  {Starobinsky}}\ and\ \bibinfo {author} {\bibfnamefont {J.}~\bibnamefont
  {Yokoyama}},\ }\href {\doibase 10.1103/PhysRevD.50.6357} {\bibfield
  {journal} {\bibinfo  {journal} {Phys.Rev.}\ }\textbf {\bibinfo {volume}
  {D50}},\ \bibinfo {pages} {6357} (\bibinfo {year} {1994})},\ \Eprint
  {http://arxiv.org/abs/astro-ph/9407016} {arXiv:astro-ph/9407016 [astro-ph]}
  \BibitemShut {NoStop}%
\bibitem [{\citenamefont {Greene}\ and\ \citenamefont
  {Kofman}(2000)}]{Greene:2000ew}%
  \BibitemOpen
  \bibfield  {author} {\bibinfo {author} {\bibfnamefont {P.~B.}\ \bibnamefont
  {Greene}}\ and\ \bibinfo {author} {\bibfnamefont {L.}~\bibnamefont
  {Kofman}},\ }\href {\doibase 10.1103/PhysRevD.62.123516} {\bibfield
  {journal} {\bibinfo  {journal} {Phys.Rev.}\ }\textbf {\bibinfo {volume}
  {D62}},\ \bibinfo {pages} {123516} (\bibinfo {year} {2000})},\ \Eprint
  {http://arxiv.org/abs/hep-ph/0003018} {arXiv:hep-ph/0003018 [hep-ph]}
  \BibitemShut {NoStop}%
\bibitem [{\citenamefont {Peloso}\ and\ \citenamefont
  {Sorbo}(2000)}]{Peloso:2000hy}%
  \BibitemOpen
  \bibfield  {author} {\bibinfo {author} {\bibfnamefont {M.}~\bibnamefont
  {Peloso}}\ and\ \bibinfo {author} {\bibfnamefont {L.}~\bibnamefont {Sorbo}},\
  }\href {\doibase 10.1088/1126-6708/2000/05/016} {\bibfield  {journal}
  {\bibinfo  {journal} {JHEP}\ }\textbf {\bibinfo {volume} {0005}},\ \bibinfo
  {pages} {016} (\bibinfo {year} {2000})},\ \Eprint
  {http://arxiv.org/abs/hep-ph/0003045} {arXiv:hep-ph/0003045 [hep-ph]}
  \BibitemShut {NoStop}%
\bibitem [{Note2()}]{Note2}%
  \BibitemOpen
  \bibinfo {note} {Eq. \protect \textup {\hbox {\mathsurround \z@ \protect
  \normalfont (\ignorespaces \ref {inflationnumber}\unskip \@@italiccorr )}}
  has been derived by assuming $m / H \not =0$, and taking the limit $k/(aH)
  \to 0$. At $m = 0$, the Hamiltonian is always diagonal, and the Bogoliubov
  transformation is trivial.}\BibitemShut {Stop}%
\bibitem [{\citenamefont {Kuzmin}\ \emph {et~al.}(1985)\citenamefont {Kuzmin},
  \citenamefont {Rubakov},\ and\ \citenamefont {Shaposhnikov}}]{Kuzmin:1985mm}%
  \BibitemOpen
  \bibfield  {author} {\bibinfo {author} {\bibfnamefont {V.}~\bibnamefont
  {Kuzmin}}, \bibinfo {author} {\bibfnamefont {V.}~\bibnamefont {Rubakov}}, \
  and\ \bibinfo {author} {\bibfnamefont {M.}~\bibnamefont {Shaposhnikov}},\
  }\href {\doibase 10.1016/0370-2693(85)91028-7} {\bibfield  {journal}
  {\bibinfo  {journal} {Phys.Lett.}\ }\textbf {\bibinfo {volume} {B155}},\
  \bibinfo {pages} {36} (\bibinfo {year} {1985})}\BibitemShut {NoStop}%
\bibitem [{\citenamefont {Chung}\ \emph {et~al.}(2015)\citenamefont {Chung},
  \citenamefont {Yoo},\ and\ \citenamefont {Zhou}}]{Chung:2013rda}%
  \BibitemOpen
  \bibfield  {author} {\bibinfo {author} {\bibfnamefont {D.~J.~H.}\
  \bibnamefont {Chung}}, \bibinfo {author} {\bibfnamefont {H.}~\bibnamefont
  {Yoo}}, \ and\ \bibinfo {author} {\bibfnamefont {P.}~\bibnamefont {Zhou}},\
  }\href {\doibase 10.1103/PhysRevD.91.043516} {\bibfield  {journal} {\bibinfo
  {journal} {Phys. Rev.}\ }\textbf {\bibinfo {volume} {D91}},\ \bibinfo {pages}
  {043516} (\bibinfo {year} {2015})},\ \Eprint {http://arxiv.org/abs/1306.1966}
  {arXiv:1306.1966 [astro-ph.CO]} \BibitemShut {NoStop}%
\end{thebibliography}%

\end{document}